\documentclass[12pt]{article}
\usepackage{graphicx,bm}

\def\k{{\bm k}}
\def\u{{\bm u}}
\def\v{{\bm v}}
\def\w{{\bm w}}
\def\x{{\bm x}}
\def\n{{\bm n}}
\def\Re{{\rm Re}}
\def\bomega{{\bm \omega}}
\def\etal{{\it et al.}}
\def\norm#1{\left|\mkern-2mu\left|#1\right|\mkern-2mu\right|}
\def\norm#1{\left|\mkern-2mu\left|#1\right|\mkern-2mu\right|}

\begin{document}

\begin{center}{{\large \bf The number of degrees of freedom of 
three-dimensional Navier--Stokes turbulence}\\~\\
Chuong V. Tran\footnote{chuong@mcs.st-and.ac.uk}\\
School of Mathematics and Statistics, University of St Andrews\\
St Andrews KY16 9SS, United Kingdom}
\end{center}
\date{\today}

\baselineskip=25pt

\centerline{\bf Abstract}
In Kolmogorov's phenomenological theory of turbulence, the energy spectrum 
in the inertial range scales with the wave number $k$ as $k^{-5/3}$ and 
extends up to a dissipation wave number $k_\nu$, which is given in 
terms of the energy dissipation rate $\epsilon$ and viscosity $\nu$ 
by $k_\nu\propto(\epsilon/\nu^3)^{1/4}$. This result leads to Landau's 
heuristic estimate for the number of degrees of freedom that scales 
as $\Re^{9/4}$, where $\Re$ is the Reynolds number. Here we consider the
possibility of establishing a quantitative basis for these results 
from first principles. In particular, we examine the extent to which
they can be derived from the three-dimensional Navier--Stokes system,
making use of Kolmogorov's hypothesis of finite and viscosity-independent
energy dissipation only. It is found that the Taylor microscale wave 
number $k_T$ (a close cousin of $k_\nu$) can be expressed in the form 
$k_T \le CU/\nu = (CU/\norm{\u})^{1/2}(\epsilon/\nu^3)^{1/4}$. Here $U$ 
and $\norm{\u}$ are, respectively, a ``microscale'' velocity and the 
root mean square velocity, and $C\le1$ is a dynamical parameter. This 
result can be seen to be in line with Kolmogorov's prediction for 
$k_\nu$. Furthermore, it is shown that the minimum number of greatest 
Lyapunov exponents whose sum becomes negative does not exceed $\Re^{9/4}$, 
where $\Re$ is defined in terms of an average energy dissipation rate, 
the system length scale, and $\nu$. This result is in a remarkable 
agreement with the Landau estimate, up to a presumably slight discrepancy 
between the conventional and the present energy dissipation rates used 
in the definition of $\Re$. 

\newpage

\section{Introduction}

For the past several decades, Kolmogorov's phenomenological theory of 
turbulence$^1$ has been a starting point for a majority of theoretical, 
numerical, and experimental studies of fluid motion. A cornerstone of 
the theory is the notion of energy cascade and dissipation anomaly.
This means that no matter how small the viscosity, the large-scale 
kinetic energy that generates and sustains the turbulence can be 
transferred to the correspondingly small scales for removal by viscous 
effects. More precisely, the energy dissipation rate has been conjectured 
to remain finite and nonzero and to become independent of viscosity in 
the limit of small viscosity. This conjecture was first extended by 
Obukhov$^2$ and Corrsin$^3$ to the case of passive scalar transport and 
mixing in turbulent flows, for which the dissipation of the scalar 
variance by molecular diffusion has been thought to be diffusivity 
independent in the limit of vanishingly small diffusivity. In a similar 
manner, Batchelor$^4$ adapted Kolmogorov's theory for two-dimensional 
turbulence, by hypothesizing that the dissipation of enstrophy (half 
mean square vorticity) becomes independent of viscosity (and remains 
nonzero) in the inviscid limit. Like the energy dissipation rate in the 
Kolmogorov theory, the supposedly finite dissipation rates of the scalar 
variance and enstrophy are key parameters in the Obukhov--Corrsin and 
Batchelor theories. Since then, the concept of cascade and dissipation 
anomaly has been thought to apply to a variety of fluid systems in other 
contexts. Thus, the Kolmogorov theory of turbulence evidently has become 
one of the most influential theories in science. 

An important prediction of Kolmogorov's theory is that the energy cascading
range (known as energy inertial range) scales with the wave number $k$ as
$k^{-5/3}$ and extends up to a dissipation wave number $k_\nu$, which is
given in terms of the viscosity-independent energy dissipation rate 
$\epsilon$ and the viscosity $\nu$ by $k_\nu\propto(\epsilon/\nu^3)^{1/4}$. 
This wave number marks the end of the energy inertial range, beyond which 
viscous forces become significant and the dissipation of energy mainly 
occurs. Heuristically, if the turbulence is homogeneous and isotropic, 
consisting predominantly of vortices of volume $k_\nu^{-3}$, then there 
are $(Lk_\nu)^3\propto(L^{4/3}\epsilon^{1/3}/\nu)^{9/4}$ such vortices 
within a domain region of size $L$. This is the basis for the Landau$^5$ 
estimate $N\propto \Re^{9/4}$ for the number of degrees of freedom $N$, 
where $\Re=L^{4/3}\epsilon^{1/3}/\nu$ is the Reynolds number.
A number of authors$^{6-11}$ have attempted to address these results 
from first principles, pending a verdict on solution regularity. The
reported upper bounds for $N$ in the literature have a wide-ranging 
dependence on $\Re$, depending on how $\Re$ is defined (and to some 
extent on how $N$ is defined). Constantin \etal$^6$ (see also Foias 
\etal$^7$) showed that the number of determining modes is proportional 
to $\Re^{9/4}$, where $\epsilon$ in $\Re$ is replaced by the asymptotic 
average of the supremum of the energy dissipation rate. They also found 
that the attractor dimension scales as $\Re^3$, where $\Re$ is defined 
in terms of the maximum fluid velocity. This dimension was found by 
Gibbon and Titi$^{9}$ to scale with the presently defined $\Re$ as 
$\Re^{18/5}$. Gibbon$^{11}$ showed that the local number of degrees of 
freedom scales as $\Re^3$, where $\Re$ was defined in terms of the 
local fluid velocity. Among these results, the first one appears to 
be closest to the Landau estimate, differing from it by the definition 
of $\epsilon$, not by the power of $\Re$. 

In this study, we revisit the Kolmogorov prediction
$k_\nu\propto(\epsilon/\nu^3)^{1/4}$ and Landau estimate
$N\propto \Re^{9/4}$, by further examining the extent to which these 
can be deduced from the three-dimensional Navier--Stokes system, making 
use of Kolmogorov's hypothesis of finite and viscosity-independent energy 
dissipation only. It is found that the Taylor microscale wave number 
$k_T$, which can be thought of as a dissipation wave number, satisfies 
$k_T \le CU/\nu$. Here $C$ is a nondimensional dynamical parameter 
satisfying $C\le1$ and $U$ denotes a ``microscale'' velocity---a 
time-dependent velocity scale associated with the enstrophy production
term. This result is derived without an {\it a priori} knowledge of 
$\epsilon$. Upon invoking Kolmogorov's hypothesis, we obtain 
$k_T \le (CU/\norm{\u})^{1/2}(\epsilon/\nu^3)^{1/4}$, where $\norm{\u}$ 
is the root mean square velocity, which can be seen to be in line with 
the prediction for $k_\nu$. Furthermore, it is shown that along any 
given bounded trajectory, 
the minimum number of greatest local Lyapunov exponents whose sum becomes 
negative is bounded from above by $\Re^{9/4}$, where $\Re$ is defined in
terms of a new domain-average energy dissipation rate. This result means 
that finite-dimensional volume elements in the infinite-dimensional phase 
space (solution space) contract exponentially if their dimensions 
exceed $\Re^{9/4}$. This is also an upper bound for generalized 
dimensions, such as the box-counting or Hausdorff dimensions, of a 
nontrivial attractor (in forced turbulence) if one exists. This result 
is consistent with the Landau estimate, up to a difference between 
the usual and present average energy dissipation rates used in the 
definition of $\Re$. This difference, which is arguably slight, is the 
``extent'' to which the present analysis needs to reach for a complete 
agreement with the Landau estimate. Thus, the present result is a step 
closer to that predicted by the classical theory. 

\section{Preliminaries}

The motion of an incompressible fluid is governed by the Navier--Stokes 
equations,
\begin{eqnarray}
\label{NS}
\u_t + (\u\cdot\nabla)\u + \nabla p &=& \nu\Delta\u, \\
\nabla\cdot\u &=& 0, \nonumber
\end{eqnarray}
where $\u(\x,t)$ is the fluid velocity, $p$ is the pressure, and $\nu$ 
is the viscosity. We consider unforced dynamics for convenience. The
results can be seen to carry over to the forced case without change,
except for a minor modification to section 3, where the forcing term
would be ignored if the injected enstrophy is negligible compared with 
that produced by the vortex stretching mechanism. This is the case for 
forcing that injects energy at large scales.$^{12}$ Equation (\ref{NS}) 
is considered 
in a periodic domain $D=[0,2\pi L]^3$, and all fields are assumed to 
have zero spatial average. This allows each component of $\u$ to be 
expressible as a Fourier series in terms of $\exp\{i\k\cdot\x/L\}$. 
Here $\k=(k_1,k_2,k_3)$, where $k_1, k_2, k_3$ are integers not 
simultaneously zero. The incompressible velocity field $\u(\x,t)$ 
can then be represented by
\begin{eqnarray}
\label{Fourier}
\u(\x,t) &=& \sum_\k\widehat\u(\k,t)\exp\left\{\frac{i\k\cdot\x}{L}\right\},
\end{eqnarray}
where $\k\cdot\widehat\u(\k,t)=0$ for incompressibility and 
$\widehat\u(\k,t)=\widehat\u^*(-\k,t)$ for reality. In other words, the 
solution vector-valued function space (phase space) can be spanned by 
the infinite basis $\{\exp[i\k\cdot\x/L]\}_\k$. Throughout this study,
we assume strong solutions up to the time under consideration. The
possibility of subsequent development of singularities is not an issue.
This allows us to make use of usual assumptions within the realm of 
the classical theory, such as bounded velocity and vorticity. 

The solution space is equipped with the scalar product 
$\langle\u\cdot\v\rangle$ and energy norm $\norm{\u}$ given, 
respectively, by
\begin{eqnarray}
\langle\u\cdot\v\rangle &=& 
(2\pi L)^{-3}\left(\int_D\u\cdot\v\,d\x\right)
\end{eqnarray}
and
\begin{eqnarray}
\norm{\u} &=& \langle|\u|^2\rangle^{1/2}.
\end{eqnarray}
The advection term $(\u\cdot\nabla)\u$ conserves the kinetic energy 
$\norm{\u}^2/2$. More generally, one has by integration by parts the
identity
\begin{eqnarray}
\label{conservation}
\langle\v\cdot(\u\cdot\nabla)\w\rangle &=& 
-\langle\w\cdot(\u\cdot\nabla)\v\rangle,
\end{eqnarray}
for all admissible $\v$ and $\w$. Also by integration by parts followed
by the Cauchy--Schwarz inequality, we have
\begin{eqnarray}
\label{csch}
\norm{\nabla\u}^2 &=& -\langle\u\cdot\Delta\u\rangle 
\le \norm{\u}\norm{\Delta\u}.
\end{eqnarray}

Consider a set of mutually orthonormal functions $\{\v_1,\v_2,\cdots,\v_n\}$, 
i.e., $\langle\v_i\cdot\v_j\rangle=\delta_{ij}$ with $\delta_{ij}$ being 
the Kronecker delta symbol. For large $n$, the number of Fourier modes
within the wave number radius $n^{1/3}/L$ is approximately $n$. Their 
(repeated) eigenvalues under $-\Delta$ are 
$k^2/L^2=(k_1^2+k_2^2+k_3^2)/L^2\le n^{2/3}/L^2$ and sum up to 
approximately $n^{5/3}/L^2$. Since these constitute the smallest 
eigenvalues of $-\Delta$, the Rayleigh--Ritz principle implies that
\begin{eqnarray}
\label{rl}
\sum_{j=1}^n\norm{\nabla\v_j}^2 &\ge& c\,\frac{n^{5/3}}{L^2},
\end{eqnarray}
where $c$ is a nondimensional constant independent of the orthonormal 
set $\{\v_1,\v_2,\cdots,\v_n\}$. 

\section{Taylor microscale and number of active modes}

In this section we consider the equation governing the evolution of 
enstrophy $\norm{\nabla\u}^2/2$, from which the enstrophy production 
and dissipation terms are compared. From this comparison during
the growing phase of the enstrophy, up to a maximum, we derive two 
expressions for the Taylor microscale wave number and compare them 
with $k_\nu$. We then rephrase the classical argument leading to the 
Landau estimate for the number of degrees of freedom.

By taking the scalar product of the momentum equation of Eq.\ (\ref{NS}) 
with $\Delta\u$ and integrating by parts the resulting time derivative 
term, we obtain
\begin{eqnarray}
\label{Z1}
\norm{\nabla\u}\frac{d}{dt}\norm{\nabla\u} 
&=& 
\langle\Delta\u\cdot(\u\cdot\nabla)\u\rangle 
- \nu\norm{\Delta\u}^2.
\end{eqnarray}
The triple-product term in Eq.\ (\ref{Z1}) is responsible for enstrophy
production, which holds the key to our understanding of turbulence. As 
it stands, this term is strongly dominated by the small scales because 
of the factors $\nabla\u$ and $\Delta\u$. Indeed, one can readily
appreciate this claim by rewriting the enstrophy production term in 
the more familiar form $\langle\bomega\cdot(\bomega\cdot\nabla)\u\rangle$, 
where $\bomega=\nabla\times\u$ is the vorticity, which clearly involves
small-scale quantities only. This means that the ``macroscale'' velocity 
field plays a largely insignificant role in the enstrophy dynamics. In 
other words, the interactions among the small scales are responsible for 
intense enstrophy production rather than the advection of small-scale
eddies and filaments by the large-scale flow. This can be true even if 
the turbulence has energetic large-scale structures superimposed on a 
sea of small-scale vortices. Motivated by this observation, we define 
a ``microscale'' velocity $U$ by
\begin{eqnarray}
\label{averagev}
U &=& \frac{|\langle(\Delta\u\cdot(\u\cdot\nabla)\u\rangle|}
{\norm{\nabla\u}\norm{\Delta\u}} =
\frac{|\langle\bomega\cdot(\bomega\cdot\nabla)\u\rangle|}
{\norm{\nabla\u}\norm{\Delta\u}}.
\end{eqnarray} 
As it stands,
$U$ can be considered as the typical velocity at the dissipation scale. 
Since $U$ is essentially the domain average of the velocity weighted by 
two functions of unity norm, it satisfies $U\le\norm{\u}_\infty$, where 
$\norm{\u}_\infty$ is the maximum fluid velocity. Given the steep energy 
spectrum of the classical theory, it is likely that $U\ll\norm{\u}_\infty$. 
It may probably be the case that $U\ll\norm{\u}$. Anyhow, with definition 
(\ref{averagev}), Eq.\ (\ref{Z1}) can be rewritten as
\begin{eqnarray}
\label{Z2}
\norm{\nabla\u}\frac{d}{dt}\norm{\nabla\u}
&\le& 
U\norm{\Delta\u}\norm{\nabla\u} 
- \nu\norm{\Delta\u}^2 \nonumber\\
&=& 
\frac{\norm{\Delta\u}^2}{\norm{\nabla\u}^2}
\left(\frac{U\norm{\nabla\u}^3}{\norm{\Delta\u}} 
- \nu\norm{\nabla\u}^2\right).
\end{eqnarray}
During the period of enstrophy increase, up to a local maximum or global 
maximum, i.e., maximum energy dissipation, Eq.\ (\ref{Z2}) implies
\begin{eqnarray}
\label{kt1}
k_T &=& \frac{\norm{\nabla\u}}{\norm{\u}} \le
\frac{U\norm{\nabla\u}^2}{\nu\norm{\u}\norm{\Delta\u}}
= \frac{CU}{\nu},
\end{eqnarray}
where $k_T$ is the Taylor microscale wave number mentioned earlier and 
$C=\norm{\nabla\u}^2/(\norm{\u}\norm{\Delta\u})$ is a nondimensional
parameter. By Eq.\ (\ref{csch}), $C$ satisfies $C\le1$, but this bound
can be highly excessive. For example, for the classical spectrum, $C$ 
tends to zero quite rapidly (as $k_\nu^{-1/3}$) in the limit 
$k_\nu\to\infty$. Note that 
without a prior knowledge of $k_\nu$, one could define $k_T$ as the 
energy dissipation wave number. Within the context of Kolmogorov's 
theory, we have $k_T\le k_\nu$; and the equality would occur for a 
$k^{-1}$ energy spectrum.

Equation (\ref{kt1}) has been derived independently of the Kolmogorov
hypothesis of finite energy dissipation $\epsilon$. This hypothesis
immediately implies that $k_T=(\epsilon/\nu)^{1/2}/\norm{\u}$. Here, 
we are interested in its dynamical consequences rather than this 
immediate implication. Together with the predicted $k^{-5/3}$ spectrum, 
the hypothesis has a profound implication on $U$, i.e., on the enstrophy 
production term. One can see from Eq.\ (\ref{Z2}) that a finite and 
viscosity-independent energy dissipation rate 
requires that $U\norm{\nabla\u}^3/\norm{\Delta\u}$ remain finite. This 
implies the scaling $U\propto k_\nu^{-1/3}$ since 
$\norm{\nabla\u}^3/\norm{\Delta\u}\propto k_\nu^{1/3}$. Thus, our 
suggestion that $U\ll\norm{\u}$ is fully justified within the framework
of the classical theory. In any case, Eq.\ (\ref{kt1}) can be rewritten 
in terms of $\epsilon$ as
\begin{eqnarray}
\label{kt2}
k_T &\le& \left(\frac{CU}{\norm{\u}}\right)^{1/2}
\left(\frac{\epsilon}{\nu^3}\right)^{1/4}.
\end{eqnarray}
This brings $k_T$ closer to $k_\nu$ in form. Note that if we invoke
the Cauchy-Schwarz inequality, i.e., replacing $C$ by its upper bound 
of unity and use $U\le\norm{\u}_\infty$, we would have 
$k_T\le(\norm{\u}_\infty/\norm{\u})^{1/2}(\epsilon/\nu^3)^{1/4}$,
which is apparently an excessive estimate.

The dissipation wave number $k_\nu$ naturally defines the number of 
degrees of freedom, being the number of dynamically active Fourier 
modes having wave numbers not exceeding $k_\nu$. This number, denoted 
by $N_c$, is approximately given by 
\begin{eqnarray}
\label{activemodes}
N_c &=& \left(\frac{k_\nu}{k_0}\right)^3 = 
\left(\frac{L^{4/3}\epsilon^{1/3}}{\nu}\right)^{9/4} = \Re^{9/4},
\end{eqnarray}
where $k_0=1/L$ is the lowest wave number. Equation (\ref{activemodes}) 
essentially rephrases Landau's estimate mentioned earlier based on the 
number of vortices having volume $k_\nu^{-3}$ within the domain $D$. 
In what follows, we show that this result largely agrees with an estimate 
for the number of degrees of freedom defined within the context of 
dynamical systems theory.

\section{Number of degrees of freedom}

This section revisits the notion of number of degrees of freedom from 
the perspective of dynamical systems theory. This number is then 
estimated for three-dimensional Navier--Stokes turbulence, using what 
is essentially equivalent to the ``trace formula,'' which was derived 
in the 1980s.$^{13,14}$ Since then, this formula has become the tool 
for virtually every study on attractor dimension of turbulence in both 
two and three dimensions. The present formulation is a simple version 
(with other advantages in addition to simplicity, see below) of the 
highly technical formulation that leads to the trace formula. It is 
an extension of a recent study by Tran and Blackbourn$^{15}$ on 
two-dimensional turbulence and is summarized in what follows. 

Chaotic dynamics are characterized by the sensitive dependence of
solutions on initial conditions, resulting in a rapid separation of 
nearby trajectories---solution ``curves''---in phase space. In a
neighborhood of a given point on a given trajectory (a given solution 
at a given time), this separation is greatest in the orthogonal 
``directions'' corresponding to the greatest local Lyapunov exponents. 
These constitute the most unstable directions of the dynamics linearized 
about the solution under consideration. In general, these directions and 
the associated exponents can change continuously along the trajectory, an 
underpinning feature of dynamical complexity. In an infinite-dimensional 
dissipative system, the number of positive local Lyapunov exponents 
along a given bounded trajectory is presumably finite, followed by a 
spectrum of negative exponents corresponding to stable orthogonal
directions. The smallest number of greatest exponents whose sum becomes 
negative (hereafter denoted by $N$) is significant as phase space 
$n$-dimensional volume elements along the trajectory contract 
exponentially for $n\ge N$. When $N$ is common for all points on an 
arbitrary bounded trajectory, volume contraction becomes universal on 
bounded sets of phase space. This number is an upper bound for the 
so-called Lyapunov or Kaplan and 
Yorke dimension$^{16,17}$ of an attractor if one exists. It is also an 
upper bound for other generalized dimensions, such as the box-counting 
and Hausdorff dimensions, of the attractor. Such an $N$ represents the 
number of degrees of freedom of the dynamical system in question, in 
the sense that its chaotic dynamics can be adequately described by an 
$N$-dimensional model. This makes sense even for cases in which no 
nontrivial attractors are known to exist. Furthermore, $N$ is well 
defined regardless of whether or not conventional Lyapunov exponents 
exist. Another advantage of the present formulation is that the problem
of global regularity of solutions of the Navier--Stokes system is not
an issue. The reason is that $N$ is determined pointwise in time, 
therefore remaining valid up to the time of solution blowup should this 
turn out to be the case.

\subsection{Local Lyapunov exponents}

Consider the linear evolution of a disturbance $\v$ to the solution 
$\u(\x,t)$ of Eq.\ (\ref{NS}) commencing from a smooth initial field 
$\u_0=\u(\x,0)$. The governing equations for $\v$ are
\begin{eqnarray}
\label{disturbance}
\v_t + (\u\cdot\nabla)\v + (\v\cdot\nabla)\u + \nabla p' &=& \nu\Delta\v,\\
\nabla\cdot\u = 0 &=& \nabla\cdot\v, \nonumber
\end{eqnarray}
where $p'$ is the perturbed pressure. By taking the scalar product of
$\v$ with Eq.\ (\ref{disturbance}) and noting that both
$\langle\v\cdot(\u\cdot\nabla)\v\rangle$ and 
$\langle\v\cdot\nabla p'\rangle$ vanish, we obtain the equation 
governing the evolution of $\norm{\v}$,
\begin{eqnarray}
\label{disturbnorm}
\norm{\v}\frac{d}{dt}\norm{\v} &=&
-\langle\v\cdot(\v\cdot\nabla)\u\rangle - \nu\norm{\nabla\v}^2.
\end{eqnarray}
It follows that
\begin{eqnarray}
\label{exponent}
\lambda = \frac{d}{dt}\ln\norm{\v} &=&
\frac{-1}{\norm{\v}^2}\left(\langle\v\cdot(\v\cdot\nabla)\u\rangle + 
\nu\norm{\nabla\v}^2\right).
\end{eqnarray} 
Here $\lambda$ is the exponential growth or decay rate of $\norm{\v}$.

The greatest local Lyapunov exponent and the corresponding most unstable 
direction can be found by maximizing $\lambda$ with respect to all 
admissible disturbances $\v$. We denote by $(\lambda_1,\v_1)$ the solution 
of this problem, where for convenience (and without loss of generality) 
$\v_1$ has been normalized, i.e., $\norm{\v_1}=1$. The second greatest 
exponent $\lambda_2$ and the corresponding second most unstable direction 
$\v_2$ orthogonal to $\v_1$ can be obtained by maximizing $\lambda$ with
respect to all disturbances $\v$ subject to the orthogonality constraint
$\langle\v\cdot\v_1\rangle=0$. Similarly, the pair of third greatest
exponent $\lambda_3$ and third most unstable direction $\v_3$ can be
obtained by solving the same maximization problem, where the admissible
disturbances $\v$ satisfy the constraint 
$\langle\v_\cdot\v_1\rangle=\langle\v_\cdot\v_2\rangle=0$. By repeating 
this procedure $n$ times, we obtain the set $\{\v_1,\v_2,\cdots,\v_n\}$ 
of mutually orthonormal functions and the corresponding ordered set of 
exponents $\lambda_1\ge\lambda_2\ge\cdots\ge\lambda_n$. These can be 
described more formally by
\begin{eqnarray}
\label{exponents}
\lambda_j &=& \max_{\norm{\v}=1}\{-\langle\v\cdot(\v\cdot\nabla)\u\rangle -
\nu\norm{\nabla\v}^2\} \nonumber\\ 
&=& - \langle\v_j\cdot(\v_j\cdot\nabla)\u\rangle -
\nu\norm{\nabla\v_j}^2
\end{eqnarray} 
for $1 \le j \le n$, where the maximization is subject to the constraint 
$\langle\v\cdot\v_i\rangle=0$ for $i<j$. These exponents provide a 
complete picture of solution stability with respect to disturbances. In 
passing, it is worth mentioning that in the maximization problem, the 
solutions $\v_j$ arise as compromises between the triple-product and 
viscous dissipation terms. That means that they do not necessarily 
maximize the former. It would be interesting to have a knowledge of the 
maximizers $(\lambda'_j,\v'_j)$ of the triple-product term alone. A 
comparison between the two Lyapunov spectra $\lambda_j$ and $\lambda'_j$ 
and between $\norm{\nabla\v_j}$ and $\norm{\nabla\v'_j}$, conceivably 
by numerical methods, could provide some invaluable dynamical insights.

\subsection{Upper bounds for the number of degrees of freedom}

We now calculate the number of degrees of freedom $N$ described earlier
by minimizing $n$ such that the sum $\sum_{j=1}^n\lambda_j$ is negative. 
By Eq.\ (\ref{exponents}), we have
\begin{eqnarray}
\label{sum}
\sum_{j=1}^n\lambda_j &=& 
- \sum_{j=1}^n\left(\langle\v_j\cdot(\v_j\cdot\nabla)\u\rangle + 
\nu\norm{\nabla\v_j}^2\right) \nonumber\\
&=& 
\sum_{j=1}^n\left(\langle\u\cdot(\v_j\cdot\nabla)\v_j\rangle - 
\nu\norm{\nabla\v_j}^2\right),
\end{eqnarray} 
where Eq.\ (\ref{conservation}) has been used. Similar to the definition 
of $U$ in the preceding section, we define two average quantities $U'$ 
and $\Omega$, respectively, by
\begin{eqnarray}
\label{averagev'}
U' &=& \frac{1}{\left(n\sum_{j=1}^n\norm{\nabla\v_j}^2\right)^{1/2}}
\left|\sum_{j=1}^n\langle\u\cdot(\v_j\cdot\nabla)\v_j\rangle\right|
\end{eqnarray} 
and
\begin{eqnarray}
\label{averagez}
\Omega &=& \frac{1}{n}\left|\sum_{j=1}^n\langle\v_j\cdot
(\v_j\cdot\nabla)\u\rangle\right|. 
\end{eqnarray} 
Like $U$ in Eq.\ (\ref{averagev}), $U'\le\norm{\u}_\infty$ represents 
a small-scale velocity by virtue of its very definition. The reason 
is that the orthonormal set $\{\v_1,v_2,\cdots,\v_n\}$ consists of 
progressively smaller-scale functions $\v_j$, i.e., increasingly greater 
$\norm{\nabla\v_j}$ as the index $j$ increases. Note that a suitable
rearrangement of the set may be necessary if it is not already in that 
order. Furthermore, similar to the $j$-th eigenvalue of $-\Delta$, we 
have $\norm{\nabla\v_j}^2\propto j^{2/3}$. This implies that $U'$ is more
strongly weighted by the smaller-scale $\v_j$'s. It is, however, not known 
with precision how $U'$ and $U$ compare. Now for definition (\ref{averagez}), 
$\Omega$ can be thought of as the domain average of $|\nabla\u|$ weighted
by $\sum_{j=1}^n|\v_j|^2/n$. By definition, the inequality 
$\Omega\le\norm{\nabla\u}_\infty$ holds. On physical grounds, this bound 
can become 
excessive for large $\Re$ because intense velocity gradients are known 
to be highly concentrated in space,$^{18,19}$ effectively getting 
``moderated'' under the spatial average in the definition of $\Omega$. 
Moreover, unless the spatial distribution of a majority of $\v_j$'s is 
strongly correlated to that of $\nabla\u$ (i.e., locally peaking in the 
same small regions as $\nabla\u$), this moderation can be more 
effective than that in $\norm{\nabla\u}=\langle|\nabla\u|^2\rangle^{1/2}$. 
The reason is that $|\nabla\u|^2$ is more resistant to such moderation 
than $|\nabla\u|$ as is reflected in the fact that 
$\langle|\nabla\u|\rangle\le\norm{\nabla\u}$. Hence, even though 
$\Omega$ is undetermined, it is expected to be closer to 
$\norm{\nabla\u}$ rather than to $\norm{\nabla\u}_\infty$. This
``conjecture'' could be readily tested numerically, given the linear 
and kinematic nature of the maximization problem. On an optimistic note, 
it is worth mentioning that one cannot rule out the possibility 
$\Omega\le\norm{\nabla\u}$, even though that might seem unlikely.

Upon substituting Eqs.\ (\ref{averagev'}) and (\ref{rl}) into the second 
equation of Eq.\ (\ref{sum}) we obtain
\begin{eqnarray}
\label{sum1}
\sum_{j=1}^n\lambda_j &\le& \left(\sum_{j=1}^n\norm{\nabla\v_j}^2\right)^{1/2}
\left(U'n^{1/2}-\nu\left(\sum_{j=1}^n\norm{\nabla\v_j}^2\right)^{1/2}\right)
\nonumber\\
&\le& \left(n\sum_{j=1}^n\norm{\nabla\v_j}^2\right)^{1/2}
\left(U'-\nu\,\frac{c^{1/2}n^{1/3}}{L}\right).
\end{eqnarray} 
The condition $\sum_{j=1}^n\lambda_j\le0$ requires a straightforward
lower bound for $n$, from which we deduce the bound
\begin{eqnarray}
\label{result1}
N &\le& c^{-3/2}\left(\frac{U'L}{\nu}\right)^3 = \Re^3,
\end{eqnarray} 
where $c^{-3/2}$ has been incorporated into the newly defined Reynolds 
number. Similar results have been reported by Constantin \etal$^6$ and
Gibbon,$^{11}$ where their Reynolds numbers were defined in terms of 
$\norm{\u}_\infty$ and of the local velocity $|\u(\x,t)|$, respectively. 
The Gibbon estimate (local number of degrees of freedom) becomes the 
Constantin estimate where $|\u(\x,t)|$ peaks. 

The above estimate has made no use of the assumption of finite energy
dissipation. Now if we identify $\nu \Omega^2$ with $\epsilon$, then 
upon substituting Eqs.\ (\ref{averagez}) and (\ref{rl}) into the first 
equation of Eq.\ (\ref{sum}), we obtain 
\begin{eqnarray}
\label{sum2}
\sum_{j=1}^n\lambda_j &\le& n\Omega-\nu\sum_{j=1}^n\norm{\nabla\v_j}^2
\nonumber\\
&\le& n\left(\frac{\epsilon^{1/2}}{\nu^{1/2}}-\nu\,\frac{cn^{2/3}}{L^2}\right).
\end{eqnarray} 
The condition $\sum_{j=1}^n\lambda_j\le0$ requires a straightforward lower
bound for $n$, from which we deduce the bound
\begin{eqnarray}
\label{result2}
N &\le& c^{-3/2}\left(\frac{L^{4/3}\epsilon^{1/3}}{\nu}\right)^{9/4} 
= \Re^{9/4},
\end{eqnarray} 
where again $c^{-3/2}$ has been incorporated into the Reynolds number 
$\Re$. This result differs from the Landau estimate by the use of
$\Omega$ instead of $\norm{\nabla\u}$ in the definition of $\Re$. This
difference can be slight as argued above. In the classical picture of 
homogeneous turbulence, there would hardly be any distinction between 
$\norm{\nabla\u}$ and $\Omega$. 

In passing, it is worth mentioning that the sum of the triple-product 
terms in the first equation of Eq.\ (\ref{sum}) is quite susceptible 
to sophisticated (and potentially excessive) estimates, which we have 
thus far deliberately avoided. Consider, for example, the Lieb--Thirring 
inequality$^{20,21}$ concerning the orthonormal set 
$\{\v_1,\v_2,\cdots,\v_n\}$, 
\begin{eqnarray}
\label{LT}
\norm{\sum_{j=1}^n|\v_j|^2} &\le& 
c'L^{3/2}\left(\sum_{j=1}^n\norm{\nabla\v_j}^2\right)^{3/4},
\end{eqnarray}
where $c'$ is a nondimensional constant independent of the set 
$\{\v_1,\v_2,\cdots,\v_n\}$. By applying Eq.\ (\ref{LT}) 
to the first equation of Eq.\ (\ref{sum}), via the intermediate step
\begin{eqnarray}
\label{Sch}
\sum_{j=1}^n\langle\v_j\cdot(\v_j\cdot\nabla)\u\rangle &\le&
\norm{\sum_{j=1}^n|\v_j|^2}\norm{\nabla\u},
\end{eqnarray} 
we would arrive at 
\begin{eqnarray}
\label{sum3}
\sum_{j=1}^n\lambda_j &\le& 
c'L^{3/2}\left(\sum_{j=1}^n\norm{\nabla\v_j}^2\right)^{3/4}\norm{\nabla\u}
-\nu\sum_{j=1}^n\norm{\nabla\v_j}^2
\nonumber\\
&\le& \left(\sum_{j=1}^n\norm{\nabla\v_j}^2\right)^{3/4}
\left(c'L^{3/2}\,\frac{\epsilon^{1/2}}{\nu^{1/2}} -
\nu\,\frac{c^{1/4}n^{5/12}}{L^{1/2}}\right),
\end{eqnarray} 
where Eq.\ (\ref{rl}) has been used in the second step. It follows that
\begin{eqnarray}
\label{result3}
N &\le& \left(\frac{c'^4}{c}\right)^{3/5}\left(\frac{L^{4/3}\epsilon^{1/3}}
{\nu}\right)^{18/5} = \Re^{18/5},
\end{eqnarray} 
where the constant prefactor has been absorbed into $\Re$, which is 
now defined in terms of $\norm{\nabla\u}$ instead of $\Omega$. The price 
for this is the scaling $\Re^{18/5}$ instead of $\Re^{9/4}$ for $N$. 
This result (given in a quite different form) was derived by Gibbon 
and Titi$^{9}$ as an upper bound for the attractor dimension.

\subsection{Discussion}

In two-dimensional turbulence, $N$ has been found to satisfy$^{15}$
\begin{eqnarray}
\label{2d}
N &\le& C'\Re(1+\ln \Re)^{1/3},
\end{eqnarray} 
where $C'$ is an absolute constant and the Reynolds number $\Re$ is defined
in terms of the materially conserved vorticity, the domain size, and $\nu$.
Apart from the difference in the level of rigor in the definition of $\Re$, 
there is a sharp contrast between the nearly linear scaling of $N$ with 
$\Re$ in two-dimensional turbulence and the highly superlinear scaling 
of $N$ with $\Re$ in the present case. This is due to fundamental 
differences between the two cases. We discuss two most apparent 
discrepancies in what follows. 

One of these is due to the dimension of the physical space and is easy
to recognize. Given $n$ Fourier modes of lowest wave numbers, the
sum of their eigenvalues under $-\Delta$---a collective measure of 
viscous dissipation strength---are $\propto n^{5/3}$ and $\propto n^2$, 
in three and two dimensions, respectively. This means that for the same
Reynolds number, three-dimensional turbulence is expected to have more 
dynamically active modes than its two-dimensional counterpart. This makes 
an intuitively obvious contribution to the difference between Eqs.\ 
(\ref{result2}) and (\ref{2d}).

The other contributing factor can be attributed to the discrepancy in 
the ``effective degree'' of nonlinearity of the small-scale dynamics of
the Navier--Stokes equations in these cases. In three dimensions, the 
dynamics are highly nonlinear, effectively quadratic. In principle, the 
vortex stretching term $(\bomega\cdot\nabla)\u$ can give rise to an 
explosive vorticity growth.$^{22,23}$ On the contrary, the two-dimensional 
Navier--Stokes system is effectively nearly linear, rendering far less 
intense dynamics of the small scales---a widely recognized fact. One 
can readily appreciate this claim by a quick inspection of the equation 
governing the vorticity gradient $\nabla\omega$,
\begin{eqnarray}
\label{gradient}
\nabla\omega_t + (\u\cdot\nabla)\nabla\omega &=&
\omega \n\times\nabla\omega - (\nabla\omega\cdot\nabla)\u +
\nu\Delta\nabla\omega,\\
\nabla\cdot\u &=& 0, \nonumber
\end{eqnarray}  
where $\n$ is the normal to the fluid domain. In Eq.\ (\ref{gradient}), 
the sole effect of the first term on the right-hand side is to rotate 
$\nabla\omega$ without changing its magnitude, and the second term alone 
is responsible for vorticity gradient amplification. By ignoring the
viscous term for convenience, we can deduce from Eq.\ (\ref{gradient}) 
the equation
\begin{eqnarray}
\label{gradient1}
|\nabla\omega|_t + \u\cdot\nabla|\nabla\omega| &=& -\frac{\nabla\omega}
{|\nabla\omega|}\cdot(\nabla\omega\cdot\nabla)\u \le |\nabla\u||\nabla\omega|. 
\end{eqnarray}  
This means that following the fluid motion, $|\nabla\omega|$ can grow no
more rapidly than exponentially in time at the instantaneous rate 
$|\nabla\u|$. 
Now, since vorticity is conserved in the inviscid dynamics, $|\nabla\u|$ 
is relatively well behaved because $\norm{\nabla\u}=\norm{\omega}$ is 
conserved. Indeed, numerical evidence shows that for an initial vorticity 
reservoir at large scales, $|\nabla\u|$ remains largely unchanged up to 
and beyond the instance of peak enstrophy dissipation.$^{24}$ More 
precisely, the ratio of the irrotational strain to $\norm{\omega}$, 
initially at $\approx 2$, has been found to remain within the range 
$[2,3.5]$ throughout the said period, during which $\norm{\nabla\omega}$
grows approximately exponentially by several orders of magnitude. Hence, 
the vorticity gradient stretching term can be said to be marginally 
nonlinear, if it is to be considered {\it nonlinear} at all. The same 
remark can be made about a broad family of fluid systems in the 
geophysical context. For this case, the gradient $\nabla q$ of the 
materially conserved potential vorticity $q$ is governed by Eqs.\ 
(\ref{gradient}) and (\ref{gradient1}), with $q$ replacing $\omega$. 
For this family, the velocity gradient $\nabla\u$ is also well behaved. 
In fact, it is presumably better behaved than its counterpart in 
two-dimensional Navier--Stokes turbulence because 
$\norm{\nabla\u}<\norm{q}$. Thus, the small-scale dynamics of this 
family are effectively marginally nonlinear. 

The enstrophy production in three-dimensional turbulence is a fundamental
problem in fluid mechanics and has always been a centre of attraction for 
the turbulence community.$^{25-34}$ This is a formidable problem, 
being virtually intractable as we have come to realize. In the limit of 
large Reynolds number, analytic and dynamically independent upper bounds 
for $\langle\Delta\u\cdot(\u\cdot\nabla)\u\rangle$ tend to become so 
excessive that they render no practical value. The reason behind these 
excessive estimates is that in order to bound the norm of a quantity, 
say $\nabla\u$, one usually resorts to norms of its derivatives, such
as $\Delta\u$. As a result, when $\norm{\nabla\u}\to\infty$, its upper 
bounds usually diverge far more rapidly. Known inequalities 
applicable to the enstrophy production term invariably reduce to 
the form ``$1\le\infty$'' (or equivalently ``$0\le1$'') as 
$\norm{\nabla\u}\to\infty$. An example is the Cauchy--Schwarz inequality 
(\ref{csch}), which was given in section 2 and briefly discussed in 
section 3. There, we bypassed this inequality by introducing the dynamical
parameter $C$. Another example is the Agmond$^{35,36}$ inequality
$\norm{\u}_\infty\le c''L^{3/2}\norm{\nabla\u}^{1/2}\norm{\Delta\u}^{1/2}$, 
where $c''$ is a nondimensional constant. This could be one of the most 
generous estimates in the present context. Given the excessive nature 
of the available inequalities (when applied to high-Reynolds number 
turbulence), it is desirable, if not crucial, to develop new techniques 
that could derive dynamically binding estimates from the governing 
equations.

\section{Conclusion}

This study has examined a possible route toward a quantitative basis 
in support of the Kolmogorov prediction for the viscous dissipation 
wave number $k_\nu$ in three-dimensional Navier--Stokes turbulence 
and the associated Landau estimate for its number of degrees of freedom
$N$. For $k_\nu$, we have taken an indirect approach, by estimating 
the Taylor microscale wave number $k_T$, which is a close cousin of 
$k_\nu$. It has been found that $k_T\le CU/\nu$, where $U$ is a 
``microscale'' velocity and $C\le1$ is a dynamical parameter. When 
expressed in terms of the energy dissipation rate $\epsilon$, this 
result becomes $k_T\le(CU/\norm{\u})^{1/2}(\epsilon/\nu^3)^{1/4}$, where 
$\norm{\u}$ is the root mean square velocity. The latter can be seen to 
be in line with Kolmogorov's prediction for $k_\nu$. For $N$, we have 
taken a direct approach, by deriving an upper bound for the minimum 
number of greatest local Lyapunov exponents whose sum becomes negative. 
The calculations of this bound have been carried out in a general manner, 
therefore the obtained result is universal for bounded trajectories.  
It is an upper bound for generalized dimensions, such as the box-counting
and Haussdorff dimensions, of a nontrivial attractor (for the forced case) 
if one exists. It has been found that $N$ satisfies $N\le \Re^{9/4}$, 
where $\Re$ is defined in terms of an average energy dissipation rate, 
the system length scale, and $\nu$. This result is in a remarkable 
agreement with the Landau estimate if one identifies the conventional 
energy dissipation rate $\epsilon = \nu \norm{\nabla\u}^2$ with the 
newly defined rate $\nu\Omega^2$, where $\Omega$ is effectively the
spatial average of the velocity gradient $|\nabla\u|$ weighted by the 
average of the squares of mutually orthonormal functions. Although
$\Omega$ is essentially undetermined, we have argued that it can be
close to $\norm{\nabla\u}$. This is the ``extent'' to which the present
analysis needs to reach for a complete agreement with the Landau estimate
of the number of degrees of freedom on the basis of the Kolmogorov theory.
In the classical picture of homogeneous and isotropic turbulence, there 
would be virtually no distinction between $\norm{\nabla\u}$ and $\Omega$.

\subsection*{References}

\noindent$^1$A. N. Kolmogorov, ``Local structure of turbulence in
incompressible fluid at very high Reynolds numbers,'' Dokl.\ Akad.\ 
Nauk.\ SSSR {\bf 30} 299 (1941).

\noindent$^2$A. M. Obukhov, ``The structure of the temperature field in 
a turbulent flow,'' Izv. Akad. Nauk. SSSR, Ser. Geogr. Geofiz. {\bf 13}, 
58 (1949).

\noindent$^3$S. Corrsin, ``On the spectrum of isotropic temperature fluctuation
in isotropic turbulence,'' J.\ Appl.\ Phys.\ {\bf 22}, 469 (1951).

\noindent $^4$G. K. Batchelor, ``Computation of the energy spectrum in 
homogeneous two-dimensional turbulence,'' Phys. Fluids {\bf 12}, 233 (1969).

\noindent$^{5}$L. D. Landau and E. M. Lifshitz, {\it Fluid Mechnics}
(Addison--Wesley, 1959). 

\noindent$^{6}$P. Constantin, C. Foias, O. P. Manley, and R. Temam,
``Determining modes and fractal dimension of turbulence flows,'' 
J.\ Fluid Mech.\ {\bf 150}, 427 (1985).

\noindent$^{7}$C. Foias, O. Manley, R. Temam, and R. Rosa, 
{\it Navier-Stokes equations and turbulence} 
(Cambridge University Press, 2001).

\noindent$^{8}$P. Constantin and C. Foias, {\it Navier--Stokes Equations} 
(University of Chicago Press, 1988).

\noindent$^{9}$J. D. Gibbon and E. S. Titi, ``Atttractor dimension and 
small length scale estimates for the three-dimensional Navier--Stokes
equations,'' Nonlinearity {\bf 10}, 109 (1997).

\noindent$^{10}$C. R. Doering and J. D. Gibbon,
{\it Applied analysis of the Navier-Stokes equations} 
(Cambridge University Press, 1995).

\noindent$^{11}$J. D. Gibbon, ``Estimating intermittency in three-dimensional
Navier--Stokes turbulence,'' J.\ Fluid Mech.\ {\bf 625}, 125 (2009).

\noindent$^{12}$B. Galanti and A. Tsinober, ``Self-amplification of the
field of velocity derivatives in quasi-isotropic turbulence,'' Phys.\ 
Fluids {\bf 12}, 3097 (2000).

\noindent$^{13}$A. V. Babin and M. I. Vishik, ``Attractors of partial 
differential equations and estimate of their dimensions,'' Russ.\ Math.\
Surv.\ {\bf 38}, 151 (1983).

\noindent$^{14}$P. Constantin, C. Foias, and R. Temam, ``Attractors
representing turbulence flows,'' Mem.\ Am.\ Math.\ Soc.\ {\bf 53}, 1 (1985).

\noindent$^{15}$C. V. Tran and L. Blackbourn, ``Number of degrees of freedom
of two-dimensional turbulence,'' Phys.\ Rev.\ E {\bf 79}, 056308 (2009).

\noindent$^{16}$ J.\ Kaplan and J.\ Yorke, {\it Functional Differential 
Equations and Approximation of Fixed Points} (Springer, New York, 1979). 

\noindent$^{17}$ J. D.\ Farmer, ``Chaotic attractors of an 
infinite-dimensional dynamical system,'' Physica D {\bf 4}, 366 (1982).

\noindent$^{18}$T. Ishihara, Y. Kaneda, M. Yokokawa, K. Itakura, and A. Uno,
``Small-scale statistics in high-resolution direct numerical simulation 
of turbulence: Reynolds number dependence of one-point velocity gradient 
statistics,'' J.\ Fluid Mech.\ {\bf 592}, 335 (2007).

\noindent$^{19}$A. Tsinober, ``Is concentrated vorticity that important?'' 
Eur.\ J.\ Mech.\ B {\bf 17}, 421 (1998).

\noindent$^{20}$E. Lieb and W. Thirring, ``Inequalities for the moments of 
the eigenvalues of the Schr\"odinger Hamiltonian and their relation to 
Sobolev inequalties,'' Studies in Mathematical Physics (Princeton University
Press, 1976), pp. 269--303.  

\noindent$^{21}$A. A. Ilyin, ``Lieb--Thirring inequalities on the 
$n$-sphere and on the plane, and some applications,'' Proc.\ London Math.\
Soc.\ {\bf 67}, 159 (1993).

\noindent$^{22}$R. M. Kerr, ``Evidence of a singularity of the
three-dimensional, incompressible Euler equations,'' Phys.\ 
Fluids A {\bf 5}, 1725 (1993).

\noindent$^{23}$R. M. Kerr, ``Velocity and scaling of collapsing Euler
vortices,'' Phys.\ Fluids {\bf 17}, 075103 (2005).

\noindent$^{24}$D. G. Dritschel, C. V. Tran, and R. K. Scott, ``Revisiting
Batchelor's theory of two-dimensional turbulence,'' J. Fluid Mech. {\bf 591},
379 (2007).

\noindent$^{25}$P. A. Davidson, {\it Turbulence: An Introduction for 
Scientists and Engineers} (Oxford University Press, 2004).

\noindent$^{26}$C. R. Doering and C. Foias, ``Energy dissipation in 
body-forced turbulence,'' J.\ Fluid Mech.\ {\bf 467}, 289 (2002).

\noindent$^{27}$C. R. Doering and E. S. Titi, ``Exponential decay rate 
of the power spectrum for solutions of the Navier--Stokes equations,'' 
Phys.\ Fluids {\bf 7}, 1384 (1995).

\noindent$^{28}$U. Frisch, {\it Turbulence: The legacy of A. N. Kolmogorov} 
(Cambridge University Press, 1995).

\noindent$^{29}$S. Goto and J. C. Vassilicos, ``The dissipation rate 
coefficient of turbulence is not universal and depends on the internal
stagnation point structure,'' Phys.\ Fluids {\bf 21}, 035104 (2009).

\noindent$^{30}$T. Ishihara, T. Gotoh, and Y. Kaneda, ``Study of 
high-Reynolds number isotropic turbulence by direct numerical simulation,'' 
Ann. Rev. Fluid Mech. {\bf 41}, 165 (2009).

\noindent$^{31}$Y. Li, L. Chivillard, G. Eyink, and C. Meneveau, ``Matrix 
exponential-based closures for the turbulent subgrid-scale stress tensor,'' 
Phys.\ Rev. E {\bf 79}, 016305 (2009).

\noindent$^{32}$C. Meneveau and K. Sreenivasan, ``The multifractal nature
of turbulence energy dissipation,'' J.\ Fluid Mech.\ {\bf 224}, 429 (1991).

\noindent$^{33}$H. K. Moffatt, S. Kida, and K. Ohkitani, ``Stretched
vortices---the sinews of turbulence; large-Reynolds-number asymptotics,''
J.\ Fluid Mech.\ {\bf 259}, 241 (1994).

\noindent$^{34}$K. Ohkitani and P. Constantin, ``Numerical study on the 
Eulerian--Lagrangian analysis of Navier--stokes turbulence,'' 
Phys.\ Fluids {\bf 20}, 075102 (2008).

\noindent$^{35}$S. Agmond, {\it Lectures on Elliptic Boundary Value Problems} 
(Mathematical Studies, Van Nostrand, New York, 1965).

\noindent$^{36}$R. Temam, {\it Infinite-Dimensional Dynamical Systems in
Mechanics and Physics,} 2nd ed.\ (Springer--Verlag, New York, 1997).

\end{document}